\documentclass[useAMS,usenatbib]{mn2e}
\usepackage{graphicx}

\def\swift{{\it Swift~}}

\title[New Lag-Luminosity relation]
{Redshift Dependent Lag-Luminosity Relation in 565 BASTE  Gamma Ray Bursts}
\author[R. Tsutsui et al.]
{Ryo Tsutsui$^{1}$\thanks{E-mail: tsutsui@tap.scphys.kyoto-u.ac.jp (RT)}
, Takashi Nakamura$^{1}$, Daisuke Yonetoku$^{2}$,Toshio Murakami$^{2}$,
\newauthor Sachiko Tanabe$^{2}$,  and Yoshiki Kodama$^{2}$ \\
$^{1}$Department of Physics, Kyoto University,
Kyoto 606-8502, Japan\\
$^{2}$Department of Physics, Faculty of Science, Kanazawa University\\
}
\begin{document}


\pagerange{\pageref{firstpage}--\pageref{lastpage}} \pubyear{2002}

\maketitle

\label{firstpage}

\begin{abstract}
 We compared  redshifts $z_Y$ from Yonetoku relation and $z_{lag}$ from the lag-luminosity relation for 565 BASTE
 GRBs and were surprised to find that the correlation is
 very low.
Assuming that the luminosity is a function of both  $z_Y$ and the
 intrinsic spectral lag $\tau_{lag}$, we found a new redshift dependent lag-luminosity relation 
  as $L=7.5\times 10^{50}{\rm erg/s}(1+z)^{2.53}\tau_{lag}^{-0.282}$  with the
 correlation coefficient  of 0.77 and the chance probability of
$7.9\times 10^{-75}$.  To check the validity of this method,
 we examined the other luminosity indicator, Amati relation, 
using  $z_Y$ and the observed fluence and found  the correlation
 coefficient of 0.92 and the chance probability of $5.2\times 10^{-106}$.   Although the spectral lag is computed 
from two channels of   BATSE, our new lag-luminosity relation suggests that
 a possible lag-luminosity relation in  the \swift era
 should also depend on redshift.

\end{abstract}


\section{Introduction}
\label{sec:intro}

Several luminosity indicators have been proposed
(Fenimore \& Ramirez-Ruiz. 2000; Norris et al. 2000; Amati et al. 2002;
 Yonetoku et al. 2004; Ghirlanda et al. 2004.; Liang \& Zhang 2005;
 Firmani et al. 2006; Schaefer 2007 for a review: See also Li 2007 and
 Butler et al. 2007 for possible evolution and bias effects ).
The variability-luminosity relation is the first luminosity indicator
 which is suggested by Fenimore \& Ramirez-Ruiz (2000).
It is based on the fact that the variable GRB with high
 variability $V$ is
brighter than the smoother one with low $V$.
Norris et al. (2000) first recognized the spectral  time lag,
which is defined from two channels (25-50keV and 100-300keV)
of BATSE, as a
luminosity indicator based on six BATSE GRBs with the optically
determined  redshifts.
{}From the BeppoSAX data, Amati et al. (2002) found the correlation
between the total isotropic energy of the prompt emission and the peak
energy $E_p$ (so called  Amati relation).
Then  Yonetoku et al. (2004)
proposed $E_p$-luminosity relation(so called Yonetoku relation
).
If we use one of the  luminosity indicators under
the standard cosmological model, we can estimate the redshifts
 of GRBs whose redshifts are unknown  
(Schafer et al. 2001; Yonetoku et al. 2004; Band et al. 2004).

Since these luminosity indicators such as Yonetoku relation and
the lag-luminosity relation are independent  each other, the
redshifts derived from different indicators for the same GRB are not
necessarily the same.
In this Letter, we first examine the correlation of 
two redshifts derived from the Yonetoku relation ($z_Y$) 
and the lag-luminosity relation ($z_{lag}$) for 565 BATSE GRBs.
 In \S 2,
surprisingly we found that the correlation between $z_Y$ and
$z_{lag}$ is very low so that  we will 
re-examine the lag-luminosity relation using $z_Y$.
In \S 3 we found  a new redshift dependent 
lag-luminosity relation 
different from the original lag-luminosity relation  by
Norris et al. (2000). 
In \S 4, we discuss the origin of the new
lag-luminosity relation, using the subjet model(Ioka \& Nakamura 2000) and the
thermal model(Ryde 2004). \S 5 will be devoted to discussions.
Throughout the paper, we assume the flat-isotropic universe with 
$\Omega_m=0.30$, $\Omega_{\Lambda}=0.70$ and 
 $H_0=70$km s$^{-1}$Mpc$^{-1}$

\section{Comparison of two redshifts}
\label{sec:redshift}

Yonetoku et al. (2004) proposed $E_p$-luminosity 
relation and estimated the peak luminosity and the redshifts
 of 689 BATSE GRBs without optically determined redshifts. 
Recently,  Tanabe et al. (2007) revised the relation 
using more GRBs  and obtained 
\begin{equation}
(\frac{L}{10^{52} \rm erg \rm s^{-1}}) = 7.90 \times 10^{-5}\left[\frac{E_p^{obs}(1+z)}{1 \rm keV}\right]^{1.82}
\end{equation}
Although the power law index  is $\sim 0.2$ smaller
than Yonetoku et al. (2004), the relation is essentially 
the same so that
we adopt in this Letter this revised Yonetoku relation.

Using 6 GRBs available at that time, Norris et al. (2000) found the lag-luminosity relation as
\begin{equation}
\frac{L}{10^{51} \rm erg \rm s^{-1}} =
 2.18 \left[\frac{\tau_{lag}^{obs}   }{0.35{\rm s}(1+z)}  \right]^{-1.15}
\end{equation}
Band et al. (2004) estimated the peak luminosity and 
the redshifts using the lag-luminosity relation.

 In Yonetoku et al. (2004), at first   745 BATSE
 GRBs were sampled. 21 GRBs have $z > 12$ and 
35 have no solution satisfying Yonetoku relation  so that 
they analyzed the remaining 689 GRBs. In these 689 GRBs,
 23 GRBs have $E_{iso}/L < 1s$.
In this Letter, we  compare two redshifts $z_Y$ and $z_{lag}$ 
for the remaining 666 GRBs.
We use lags in database for 1430 BATSE burst.
We found that 621 GRBs are included in both data. 
56 GRBs have negative spectral lags
so that Eq.(2) can not be used for
these GRBs. The number of GRBs is now 565. 
Figure 1 plots $z_Y$ versus $z_{lag}$
with the solid line being $z_Y=z_{lag}$.
Surprisingly enough there are many GRBs with (1) large $z_Y$ and
small $z_{lag}$ as well as (2) small $z_Y$ and large $z_{lag}$. 
We see that  the correlation between  $z_Y$ and
$z_{lag}$ is very low.
At this point there are three possibilities;
(a) the lag-luminosity relation, (b) Yonetoku relation 
(c) both relations are  responsible for this low correlation.
\begin{figure}
\includegraphics[width=84mm]{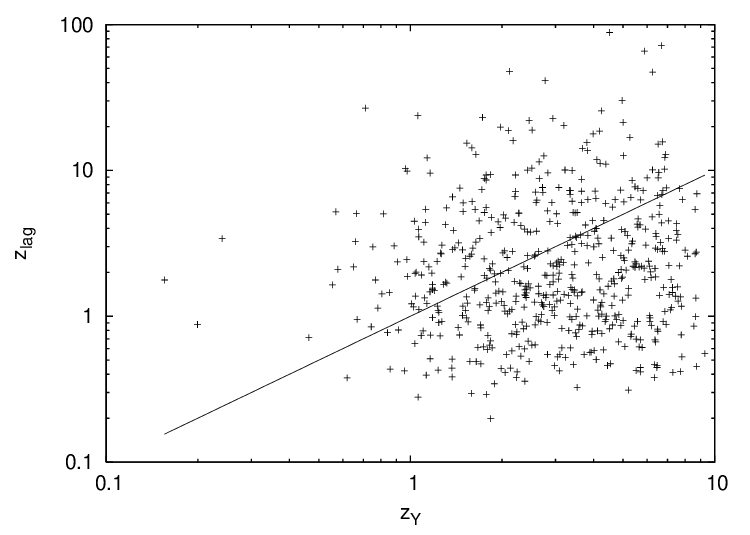}
 \vspace{0pt}
 \caption{The distribution of ($z_Y$, $z_{lag}$) for 565 BASTE GRBs
 where $z_Y$ and $z_{lag}$ are redshifts determined by Yonetoku and the lag-luminosity
 relations, respectively. The correlation is very low. The solid line is $z_Y$ =$z_{lag}$. It is expected that
($z_Y$, $z_{lag}$) distribute around the solid line.}
\label{fig1}
\end{figure}

\begin{figure}
\includegraphics[width=84mm]{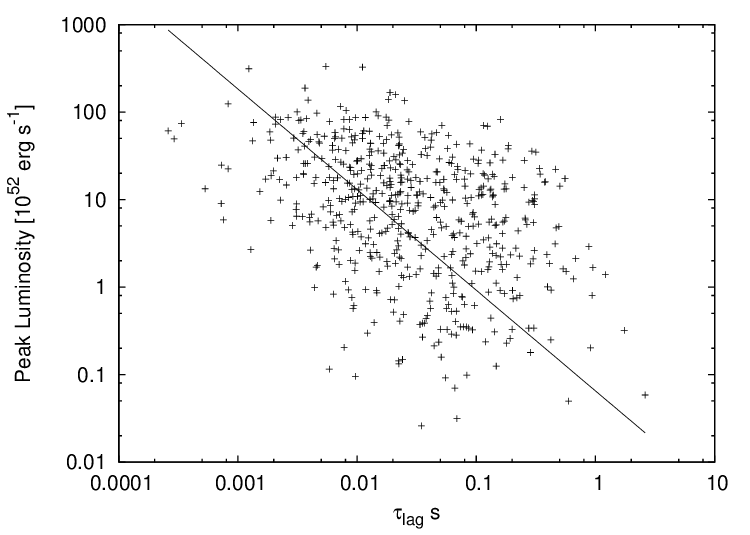}
 \vspace{0pt}
 \caption{$\tau_{lag}$ vs $L_{52}$ using $z_Y$ for 565 BASTE GRBs. 
The correlation coefficient is 0.38. The chance probability is $1.7 \times 10^{-19}$ so that the correlation coefficient is low.}
\label{fig2}
\end{figure}

We first consider the first possibility (a), since
in the revised Yonetoku relation, Tanabe et al. (2007)
examined the evolution effect as well as the observational 
selection bias and found that they are small.
In Fig. 2, we plot
 log[$\tau_{lag}$] vs log[$L_{52}$] using  $z_Y$
 where the solid line is the original
lag-luminosity relation by Norris et al. (2000). 
The correlation coefficient is 0.38  and the
chance probability  is $1.7 \times 10^{-19}$ which is rather
large considering the number of samples 565. 
The reason for this low correlation coefficient
  is the large scatter around the solid line.

\begin{figure}
\includegraphics[width=84mm]{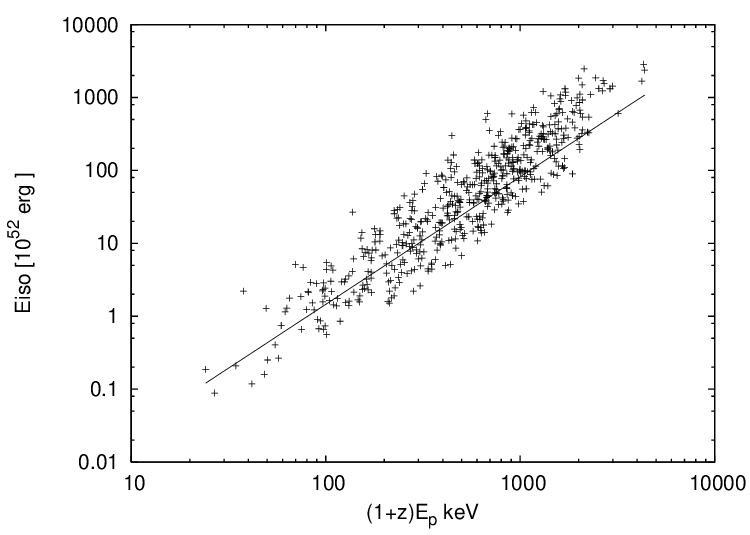}
 \vspace{0pt}
 \caption{Amati relation in 565 BATSE  GRBs. Redshifts derived from Yonetoku 
 relation  $z_Y$ is used to estimate $E_{iso}$ and $E_p(1+z)$.
The correlation 
 coefficient is 0.92. The chance probability is
 $5.2 \times 10^{-106}$
 so that the correlation is tight.
The solid line is Amati relation ( Amati 2006).}
\label{fig3}
\end{figure}
Then we like to ask what will happen if we adopt another
distance indicator such as Amati relation.
We use  $z_Y$  and plot
$E_{iso}$ and the intrinsic $E_p$ in 
 Fig. 3. The correlation 
 coefficient is 0.92. The chance probability is $5.2 \times 10^{-106}$
so that the correlation is tight. We can say that  Amati relation is
compatible with Yonetoku relation while the original lag-luminosity
relation is not so. This is also the reason why we consider
the first possibility (a). 

\section{New lag-luminosity relation}
\label{sec:lag-lumi}
\begin{figure}
\vspace{0pt}
 \begin{center}
      \begin{tabular}{cc}
      \resizebox{40mm}{!}{\includegraphics[clip]{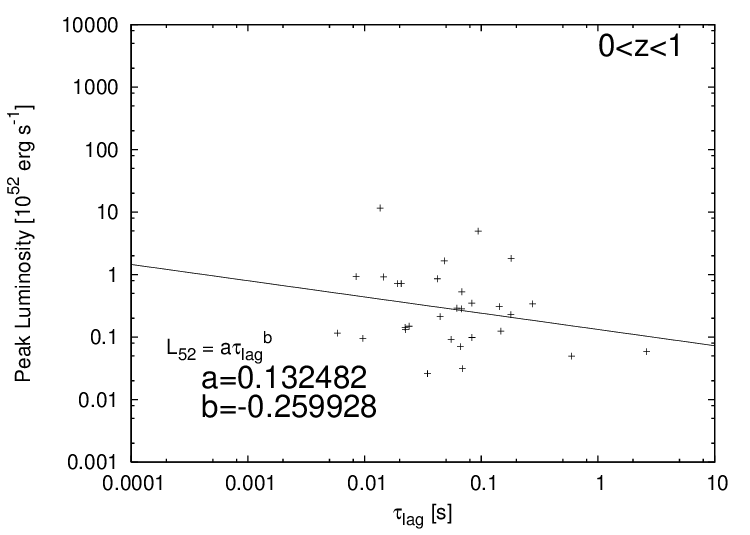}} &
      \resizebox{40mm}{!}{\includegraphics[clip]{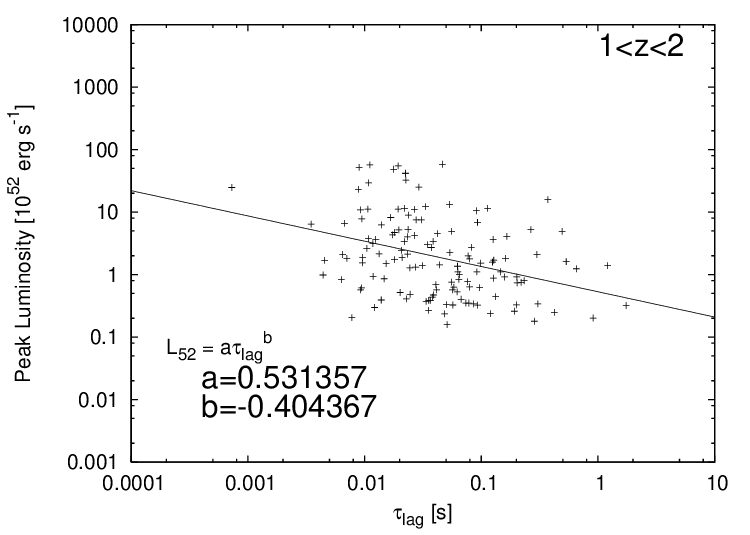}} \\
      \resizebox{40mm}{!}{\includegraphics[clip]{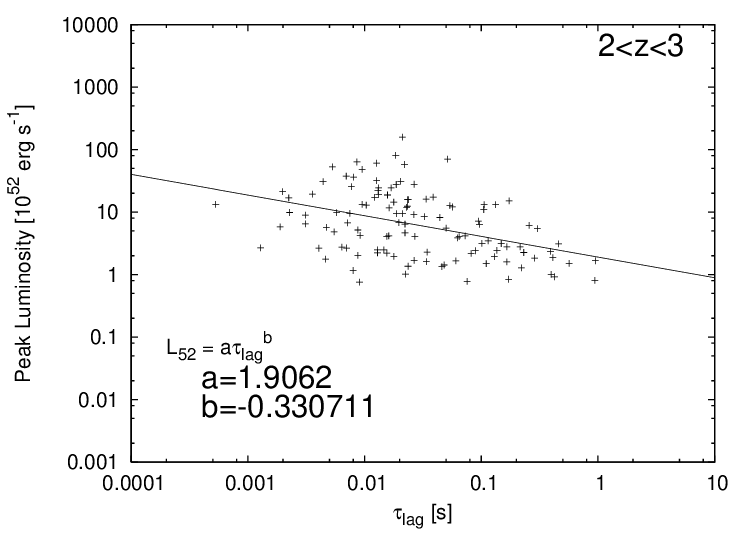}} &
      \resizebox{40mm}{!}{\includegraphics[clip]{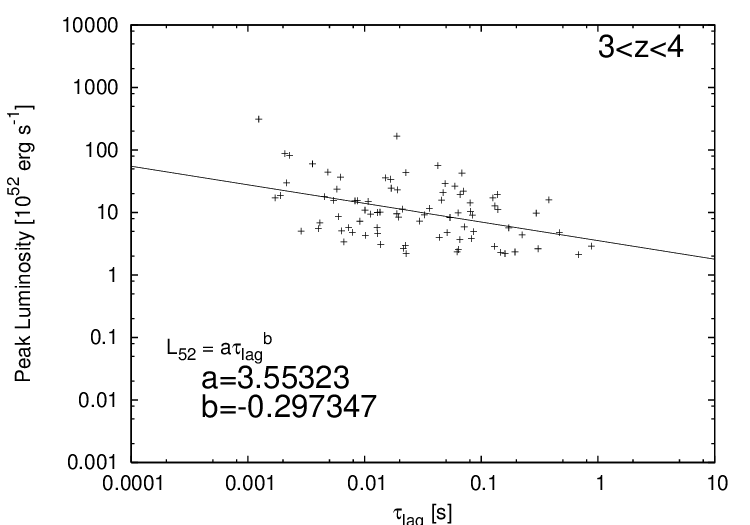}} \\
      \resizebox{40mm}{!}{\includegraphics[clip]{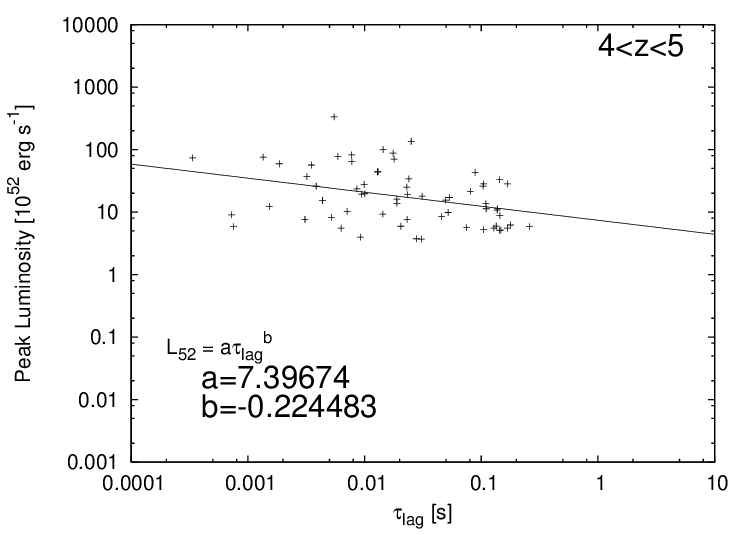}} &
      \resizebox{40mm}{!}{\includegraphics[clip]{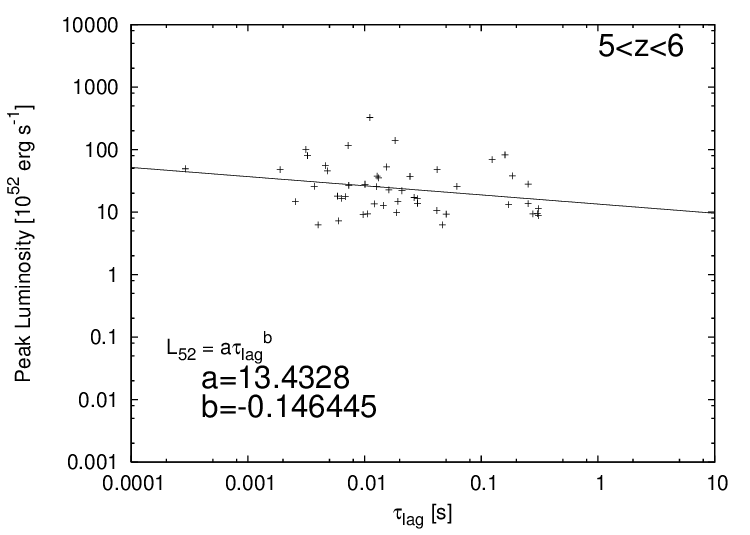}} \\
      \resizebox{40mm}{!}{\includegraphics[clip]{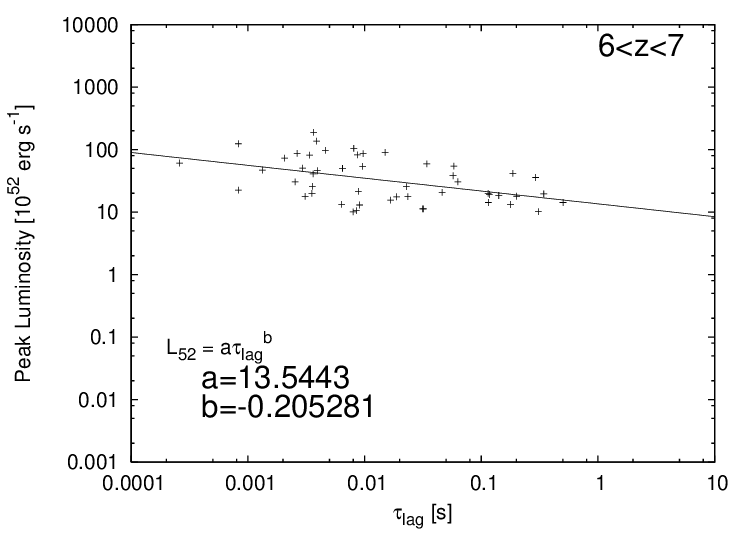}} &
      \resizebox{40mm}{!}{\includegraphics[clip]{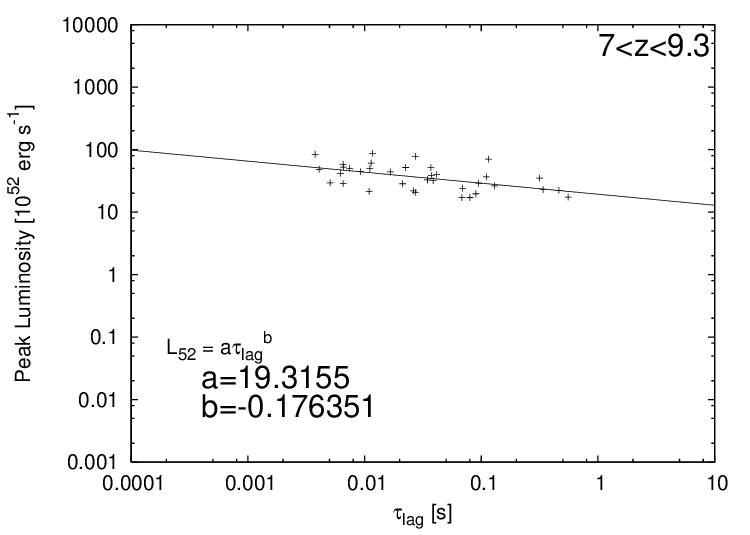}} \\
    \end{tabular}
    \caption{Lag-luminosity relations for  various redshift groups as: 0 $ \le z  < $   1 ; 1 $\le z < $  2 ; 
2 $\le z < $ 3 ;  3 $\le z < $  4 ; 4 $\le z < $ 5 ; 
5 $\le z < $ 6 ; 6 $\le z < $ 7 ; 7 $\le z < $ 9.3.devided by
  redshift. In each redshift group we tested the relation
$L=a\tau_{lag}^b$. The best fit values of $a$ and $b$ are shown in each
  figure. The solid lines are the best-fit power-law models 
  for each redshift group. We  see that $b$ is almost the same
while $a$ increases as a function of $z$.}
    \label{fig4}
 \end{center}
\end{figure}
Figure 2 shows 
that there is a large variance 
in the original lag-luminosity relation. To seek the origin of 
this variance we here ask the value of the redshift in Fig. 2.
We divide the data in Fig. 2
into  redshift groups as: 0 $ \le z  < $   1 ; 1 $\le z < $  2 ; 
2 $\le z < $ 3 ;  3 $\le z < $  4 ; 4 $\le z < $ 5 ; 
5 $\le z < $ 6 ; 6 $\le z < $ 7 ; 7 $\le z < $ 9.3.
In each redshift group, assuming the lag-luminosity relation
as $L=a\tau_{lag}^b$, we show  in Fig. 4 the least square fit by  solid lines 
 with the value of power law index $b$ and the amplitude $a$.
We see that  the  power law indices $b$ are almost the same
while the higher redshift GRBs have larger $a$. 
This suggests 
the existence of the redshift dependent effect 
in the lag-luminosity relation.
Inspired by Fig. 4, we assume that 
the luminosity is described by $L=A(1+z)^{\alpha}\tau_{lag}^{\beta}$ ,
and found that the best fit curve is given by
\begin{equation}
\log L_{52} = -1.12 +  2.53\log(1+z) -0.282\log(\tau_{lag})
\label{eq:new-lag-lumi}
\end{equation}
The standard deviation of the relation is $\sigma$ = 0.473.
 In Fig. 5 we plot
log[$0.0758(1+z)^{2.53}\tau_{lag}^{-0.282}$] versus 
log[$L_{52}$]and
 found that the  correlation coefficient
is 0.77 with the chance probability of $7.9 \times 10^{-75}$.
The correlation coefficient is much higher than the original
lag-luminosity relation.

\begin{figure}
\includegraphics[width=84mm]{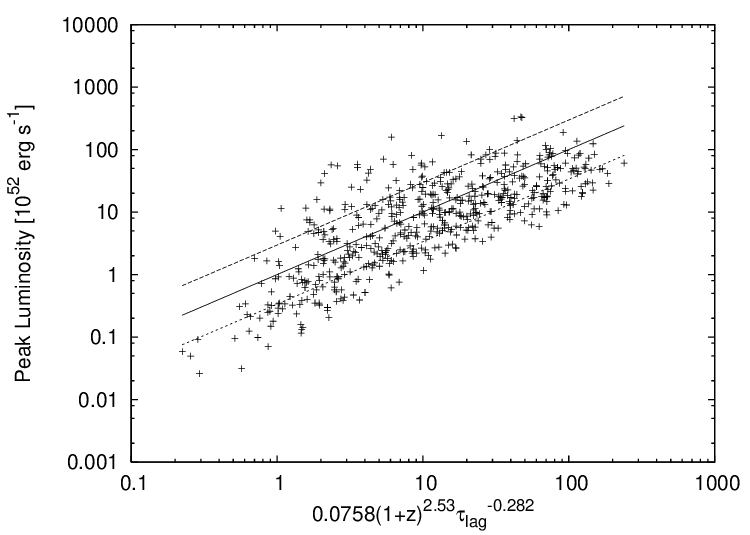}
 \vspace{0pt}
 \caption{The new redshift dependent lag-luminosity relation in
$0.0758(1+z)^{2.53}\tau_{lag}^{-0.282}$ vs $L_{52}$ plane.
 The correlation coefficient is 0.77.
 The chance probability is $7.9 \times 10^{-75}$.
The solid line is the best fitting line and
 two dashed lines are 
 1-$\sigma$ (0.47 in log10) deviation line. This
 has a lower chance probability than
the original lag-luminosity relation in Fig. 2.
}
\label{fig5}
\end{figure}

In the new lag-luminosity relation, the power law index 
for $\tau_{lag}$ is about a factor 4 smaller than that
in the original lag-luminosity relation so that
one may ask for the reason of the difference. 
We consider  the same
6 GRBs as in  Norris et al. (2000). 
Figure 6 shows the original lag-luminosity relation in
the luminosity-spectral lag plane.
 We found that the
correlation coefficient is 0.94, and the chance probability is 0.021.
In Fig. 7 we show the new lag-luminosity relation for
the same 6 GRBs in the luminosity -$0.0758(1+z)^{2.53}\tau_{lag}^{-0.282}$
plane.  We found that
the correlation coefficient is 0.90, and the chance probability is 0.027.
As for the correlation coefficients and the chance probability,
we found  no significant difference between two relations in the
original 6 GRBs. 
Therefore the new lag-luminosity relation is consistent with  6 GRBs originally used by  Norris et al. (2000).

\begin{figure}
\includegraphics[width=84mm]{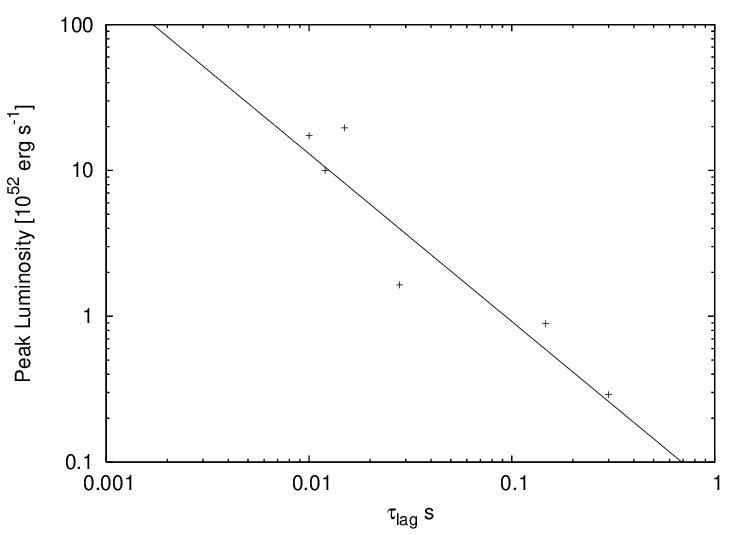}
 \vspace{0pt}
 \caption{Lag-luminosity relation for GRBs used in Norris et al. (2000). The correlation coefficient is 0.94. The chance 
 probability is 0.021. The solid line is Norris's original 
 lag-luminosity relation.
}
\label{fig6}
\end{figure}

\begin{figure}
\includegraphics[width=84mm]{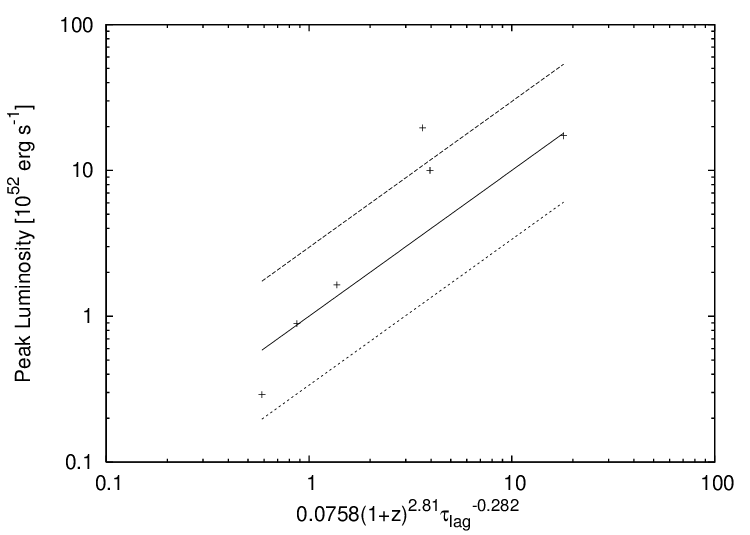}
 \vspace{0pt}
 \caption{New lag-luminosity relation for GRBs used in Norris et al. 2000.
The correlation coefficient is 0.90. The chance probability is 0.027.
The solid line represents Eq. (3).
Two dashed lines mark the 1-$\sigma$ deviation from Eq. (3).
}
\label{fig7}
\end{figure}

\section{Possible Interpretation of the new  lag-luminosity relation }
\label{sec:interpretation}

Ryde (2004) studied 5 GRBs which are consistent with a thermal blackbody 
radiation throughout their duration and found the temperature $kT$ can 
be well described by broken power law as a function of time.
Figure 11 in Ryde (2004) shows the time evolution of the temperature is
described as
\begin{equation}
 kT_{obs} \approx 100{\rm keV} \times t_{obs}^{-0.2},
\end{equation}
in the relevant early time to the spectral lag.
If the peak energy of GRB is determined by the temperature of blackbody 
spectrum (Thompson et al. 2007; Rees \& M${\acute {\rm 
e}}$sz${\acute {\rm a}}$ros 2005), we can identify $kT_{obs}$ with $E_p^{obs}$.
Assuming that $E_p$ evolves like Eq. (4) and $dt_{obs}/d E_p^{obs}$ is in proportion to
$\tau_{lag}^{obs}$ with Yonetoku relation of $L\sim E_p^2$.
 we have
\begin{equation}
 L_{52} \propto (1+z)^{1.67}\tau_{lag}^{-0.33}.
\end{equation}
This relation has the  similar value of power law index for
$\tau_{lag}$ to the new lag-luminosity relation.

Ioka \& Nakamura (2001) suggested the origin of the lag-luminosity relation is 
the viewing angle to the jet axis.
They adopted the following form of the spectrum in the comoving frame, 
which yields a spectral shape similar to the observed Band spectrum as
\begin{equation}
 f(\nu^{'}) = \left(\frac{\nu^{'}}{\nu_{0}^{'}}\right)^{1+\alpha_{B}}
\left[1+\left(\frac{\nu^{'}}{\nu_{0}^{'}}\right)^{l}\right]^{\frac{\beta_B-\alpha_B}{l}}
\end{equation}
where $l$ is a parameter which controls the smoothness of the
transition between the high energy power law and the low energy one with
$\alpha_B$ and  $\beta_B$ being the parameters 
in Band function, respectively. They adopted $l$=2 in their
 application to the original lag-luminosity relation. 
For general $l$,
we can derive the following equation
\begin{equation}
 L = \nu F_{\nu} \propto \nu^{3}{{{\delta}T_p}^{\frac{-1+\alpha_B}{l}}}
\end{equation}
where $\delta T_p$ is the spectral lag, and $\nu$ is the intrinsic frequency.
Since $\nu$ is related to  $\nu_{obs}$  fixed by 
BATSE energy channels as $\nu=(1+z)\nu_{obs}$,
  we can rewrite Eq. (7) as
\begin{equation}
 L \propto (1+z)^{3}\tau_{lag}^{-0.3}
\end{equation}
for $l=6$. This is qualitatively consistent with 
the relation in Eq. (3).

\section{Discussions}

The definition of the spectral lag depends on the redshift from
the beginning. The spectral lag is calculated from the data
of the observed two channels in \ 25-\ 50 \ keV and
 100-300 \ keV. However in GRB rest frame for $z=4$, for example,
two channels are \ 125-\ 250 \ keV and 500-1500 \ keV. 
Therefore the lower channel for $z=4$ corresponds to the higher
channel for $z=0$. If the spectral lag depends on the observed
photon energy even for $z=0$, the lag-luminosity relation should
depend on redshifts for $z\neq 0$ so that our redshift dependent new lag-luminosity
relation is not so strange. If the lag-luminosity relation does
not depend on the redshift, the spectral lag should
 not depend on the intrinsic photon energy. However the concept of the spectral lag comes from
the fact that the peak time depends on the  photon  energy.
In reality, Norris et al. (2000) showed that the lag between
channels 4($> 300$keV) and  1(25-50keV) is $\sim 2\sim 3$ times larger than that 
between channels 3(100-300keV) and 1. 

 We derived the new lag-luminosity 
relation from 565 GRBs while Norris's original relation
 was derived   from 6 GRBs. In this Letter,
the only assumption used to derive the new
 lag-luminosity relation is that Yonetoku relation is
 free from serious evolution and selection bias effects. 
The new
 lag-luminosity relation has lower chance probability than
the original lag-luminosity relation by Norris et al. (2000)
 and is compatible with 6 GRBs used in  Norris et al. (2000).

Finally we discuss redshifts determined by the
 new lag-luminosity relation.
Equation (3) can be rewritten  as 
\begin{equation}
 \frac{{d_L,_{26}}^2}{(1+z)^{2.81}} = 
  0.0758\frac{(\tau_{lag}^{obs})^{-0.282}}{4\pi F}
\label{eq:lhs}
\end{equation}
for each GRB, where $d_L$[cm] is the luminosity distance and F[erg cm$^{-2}$ s$^{-1}$] is the photon energy flux.
The left hand side  of 
 Eq. (9) as a function of $z$ begins from zero, has a maximum at
 $z\sim 4$ and then decreases. The new lag-luminosity relation
has one-$\sigma$ deviation of 0.47 in log10. Then the
right hand side of Eq. (9) changes a factor 3 so that the accuracy of redshifts is not so good. It often occurs that there is
 no solution for $z$ like in  Amati relation.

We need  tighter lag-luminosity relation to estimate 
redshifts. The spectral lag for BATSE GRBs is defined from 
two channels in BATSE. However \swift does not have such two
channels so that a new definition of a spectral lag is needed
in the \swift era. Then we may construct
the  tighter lag-luminosity relation in the \swift era using \swift GRBs with known redshifts.  Our results suggest
that such a lag-luminosity relation in the \swift era should
depend on the redshift. 

So far \swift observed $\sim$ 200 GRBs but only  $\sim 50$ GRBs
have spectroscopically determined redshifts. 
For these \swift GRBs without redshifts, 
if we can determine redshifts only from gamma ray observations, 
  redshifts might be  estimated in advance of deep follow-ups
so that possible high redshift GRBs might be selected for detailed
observations. Therefore it is urgent to find the lag-luminosity relation 
,@which does not need $E_p$, in the \swift era.

\section*{Acknowledgments}
We thank K. Ioka, K. Toma, T. Muranushi, T. Muto, and T. Kobayashi for useful comments. 
This work is supported in part by
the Grant-in-Aid from the 
Ministry of Education, Culture, Sports, Science and Technology
(MEXT) of Japan,  No.19540283,No.19047004, No.19035006(TN),
 No.18684007 (DY)
and by a Grant-in-Aid for the 21st Century COE
``Center for Diversity and Universality in Physics''
from the Ministry of Education, Culture, Sports, Science and Technology
(MEXT) of Japan

%



\label{lastpage}

\end{document}